\documentclass[aps,prd,longbibliography ,twocolumn,nofootinbib,superscriptaddress,reprint]{revtex4-2}
\usepackage[dvipsnames]{xcolor}
\usepackage{graphicx}
\usepackage{amsmath}
\usepackage{amsfonts}
\usepackage{graphicx}
\usepackage{dsfont}
\usepackage[german, english]{babel}
\usepackage{amssymb}
\usepackage{ifthen}
\usepackage{ulem}
\usepackage{color, xcolor}
\usepackage{float}
\usepackage{physics }
\usepackage[colorlinks=true,citecolor=blue,urlcolor=blue,linkcolor=blue]{hyperref}
\usepackage{bbold}
\usepackage{nccmath}

\newcommand{\secref}[1]{Sec.~\ref{#1}}
\newcommand{\appref}[1]{Appendix~\ref{#1}}
\newcommand{\figref}[1]{Fig.~\ref{#1}}
\renewcommand{\eqref}[1]{Eq.~(\ref{#1})}
\newcommand{\da}{^\dagger}
\newcommand{\ii}{\text{i}}

\begin{document}
\title{A microscopic approach to the problem of enhancement and suppression of superconductivity on twinning planes}
\author{Anton Talkachov}\
\email{anton.talkachov@gmail.com}
\affiliation{Department of Physics, KTH-Royal Institute of Technology, SE-10691, Stockholm, Sweden}
\author{Sahal Kaushik}
\affiliation{Department of Physics, KTH-Royal Institute of Technology, SE-10691, Stockholm, Sweden}
\affiliation{Nordita, Stockholm University and KTH Royal Institute of Technology, Hannes Alfv\'{e}ns v\"{a}g 12, Stockholm, SE-106 91,  Sweden}
\author{Egor Babaev}
\affiliation{Department of Physics, KTH-Royal Institute of Technology, SE-10691, Stockholm, Sweden}
\date{\today}

\begin{abstract}
Using a microscopic approach, we revisit the problem of superconducting critical temperature change in the presence of twin boundaries.
We show that both critical temperature enhancement and suppression can come purely from geometric effects. These include aspects of scattering of electrons on these crystalline defects even when the coupling constant is unchanged. 
We consider two dimensional rectangular and three dimensional body centered cubic lattices with onsite s-wave superconducting pairing, nearest and next-to-nearest neighbor hoppings.
In the considered two dimensional lattice with twin boundaries, the superconducting critical temperature associated with twinning planes is suppressed for moderate band filling and enhanced for an almost empty/filled band.
The superconducting phase diagram is more diverse for the three dimensional lattice, which is caused by the interplay of van Hove singularity, changing coordination number, and modification of distances to nearest and next-to-nearest neighbors.
    
\end{abstract}

\maketitle

\section{Introduction}
Interfaces, domain walls and twinning planes can significantly affect properties of condensed matter systems.
The crystaline defects: twin boundaries (TBs) are known to significantly affect superconducting properties, often leading to a local increase of superconducting critical temperature. 
For example, certain elemental superconductors are known to have enhanced critical temperature at TBs, while in other materials, superfluid density is suppressed at TB; the mechanism for the enhancement is an especially interesting open question.
Many previous theoretical works  focused on phenomenological description of local increase of superconducting correlations using Ginzburg-Landau  theory \cite{nabutovskij1981localized,averin1983theory,andreev1987Exotic,abrikosov1988manifestation}.
A phenomenological Ginzburg-Landau model was proposed  based on the assumption of existence of different phonons localized on a TB \cite{nabutovskij1981localized,averin1983theory}.

Besides the elemental materials, interesting
properties of twinning planes are exhibited by various more chemically complicated compounds \cite{dimos1990superconducting,BOBROV1991411,nabatame1994transport,PhysRevLett.109.137004,PhysRevX.5.031022,sigrist1996influence,walker1996josephson,feder1997twin}.
In the current paper, we address a fundamental property of TB -- the deviation of its critical temperature from a bulk value and the existence of localized superconductivity. The topic has been a subject of experimental observations for decades \cite{khlyustikov1985superconductivity,khlyustikov1987twinning,bobrov1984increase,BOBROV1991411,PhysRevLett.109.137004}.

The motivation of this study is the recent insights gained in the revision of the problem of a boundary between superconductor and vacuum and other materials \cite{samoilenka2020boundary,samoilenka2020pair,benfenati2021boundary,barkman2022elevated,talkachov2023microscopic,hainzl2023boundary,roos2023bcs,roos2023boundary}.
Although on the one hand the problem is fundamentally different, the recent works indicated that enhancement of superconducting critical temperature can be produced by very short-range effects beyond standard quasiclassical approaches.
For example, electron scattering from a boundary creates Friedel oscillations of the density of states, allowing highly inhomogeneous yet macroscopic solutions of the gap equation with higher critical temperature than bulk $T_c$.
This effect is based on electron scattering peculiarities and does not require extra pairing glue.
This motivates considering a fully microscopic model for a twinning plane without additional modifications of TB region (no introduction of additional phonon modes, changing chemical potential for mimicking depletion of the carrier density on TB \cite{feder1997twin}, etc.). 

Below we show that even without the modification of pairing glue (i.e. without change of pairing potential but merely from geometric effects), TB critical temperature can be increased or suppressed depending on band filling, TB direction (which controls coordination numbers) and the bulk critical temperature.
We also investigate TB interactions with neighbors and find parameter space regions where only nearest neighbor TBs are relevant.

The suppression of the critical temperature due to TB presence is directly captured by local probes like magnetometry measurements \cite{khlyustikov1987twinning,khlyustikov1989influence}.
Gap suppression on a TB is observed in scanning tunneling microscopy \cite{song2012suppression}.
The enhancement can be observed by both local and 
transport techniques.
They are the onset of resistivity \cite{khlyustikov1985superconductivity,khlyustikov1987twinning} and described above methods \cite{kirtley2010meissner,kalisky2011behavior,noad2016variation}.

The paper is structured as follows. In \secref{sec:the model}, we describe the Hamiltonian for the superconductivity model, develop the numerical method considered in Ref. \cite{samoilenka2020boundary}, and investigate the fact of no change in critical temperature due to TB for the models with nearest neighbor hoppings.
Next, in \secref{sec:Two dimensional rectangular lattice}, we study the superconductivity phase diagram for two dimensional rectangular lattice with next-to-nearest neighbor interactions and two TBs and explore the magnitude of critical temperature change on TB spacing.
In \secref{sec:Body centered cubic lattice}, we perform computations similar to \secref{sec:Two dimensional rectangular lattice} for body centered cubic (BCC) lattice with 112 TB.

\section{The model} \label{sec:the model}
\subsection{Superconductivity model} \label{sec:Superconductivity model}
We consider a mean-field Hubbard Hamiltonian on a lattice corresponding to a single band s-wave electron pairing without a magnetic field in the system:

\begin{equation}\label{eq:mean-field_Hamiltonian}
\begin{split}
    H_\text{MF} = & -\sum_{i,j, \sigma} h_{ij} c_{i,\sigma}\da c_{j,\sigma} \\
    & + \sum_{i} \left( \Delta_{i} c_{i,\uparrow}\da c_{i,\downarrow}\da + \Delta^{*}_{i} c_{i,\downarrow} c_{i,\uparrow}\right) + \text{const},
\end{split}
\end{equation}
where $h_{ij} = \mu \delta_{ij} + t_{ij}^{(1)} + t_{ij}^{(2)}, \quad
       \Delta_{i} = -V \langle c_{i,\downarrow} c_{i,\uparrow} \rangle$.
Here $c_{i,\sigma}\da (c_{i,\sigma})$ is the creation (annihilation) operator for an electron with spin $\sigma$ on site $i$, $\Delta_{i}$ is the superconducting order parameter, $\mu$ is the chemical potential (constant in space), $t_{ij}^{(1)}$ and $t_{ij}^{(2)}$ are the nearest and next-to-nearest neighbor hopping integrals, respectively, and $V$ is the on-site attractive ($V>0$) potential.

We are interested in the superconducting critical temperature in a system with uniform $V$, i.e. no extra pairing glue associated with TB. To calculate it, we use the linearized gap equation  (see detail of the approach in e.g. \cite{de1964boundary, samoilenka2020boundary})
\begin{equation}
\label{eq:Tc_discrete}
    \frac{1}{V} \Delta_{i} = \sum_{j} K_{ij} \Delta_{j},
\end{equation}
where $K_{ij}$ is a kernel of Bogoliubov–de Gennes equations which has form

\begin{equation} \label{eq:K_matrix}
\begin{gathered}
    K_{ij} = \sum_{m,n} F (m;n) w_m^* (i) w_n^* (i) w_m (j) w_n (j), \\
    F (m;n) = \frac{1 - f(\epsilon_m)- f(\epsilon_n)}{\epsilon_m + \epsilon_n}.
\end{gathered}
\end{equation}
Here $f(E) = (1+e^{E/{T}})^{-1}$ is the Fermi-Dirac distribution function, $w_n$ are the one-electron wave functions in the normal state corresponding to eigenenergies $\epsilon_n$, which we find numerically; $k_B = 1$.
The largest eigenvalue of the kernel $K$ gives $V^{-1}$, and the corresponding eigenvector is the order parameter distribution close to the superconducting transition.

\begin{figure}
    \begin{center}
		\includegraphics[width=0.95\columnwidth]{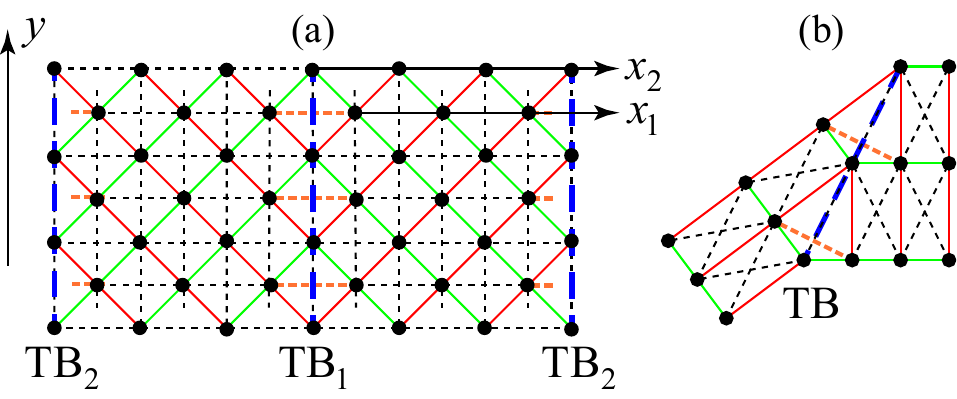}
		\caption{(a) Cartoon image of a rectangular lattice with TBs, nearest and next-to-nearest hoppings and periodic boundary conditions. The color of dashed lines corresponds to different hopping values.
                 (b)~TB region without rescaling.
                 On a TBs (indicated with blue dashed lines) the orange diagonal is shorter than the black one.}
		\label{fig:2D lattice model with NNNH and TB}
\end{center}
\end{figure}

We find superconducting critical temperature numerically.
Filling the matrix $K$ is the most time-consuming step in the algorithm. The computational complexity of the operation is $\mathcal{O} \left( N^4\right)$ where $N$ is the total number of sites in the system. However, it could be significantly reduced if the system has periodic boundary conditions (in our case, in the direction along the TB). Consider a two dimensional system with translational invariance along the $y$ direction (see \figref{fig:2D lattice model with NNNH and TB}). One could simplify the procedure by performing Fourier transform for wave functions and superconducting gap in that direction \cite{samoilenka2020boundary}:
\begin{equation} \label{eq:fourier transform}
\begin{gathered}
    w_{k_x,k_y} (i_x,i_y) = \frac{1}{\sqrt{N_y}} e^{-\ii k_y i_y} \omega_x (i_x, k_x, k_y), \\
    \Delta (i_x,i_y) = \frac{1}{\sqrt{N_y}} \sum_{k_y} e^{-\ii k_y i_y} \Tilde{\Delta} (i_x, k_y),
\end{gathered}
\end{equation}
where $k_y = \frac{2 \pi n}{N_y}$, $n \in [1;N_y]$.
Substituting the formulas to \eqref{eq:Tc_discrete} one obtains
\begin{equation}
\begin{gathered} \label{eq:linearized gap equation final}
    \frac{1}{V} \Tilde{\Delta} (i_x, k_y) = \sum_{j_x} K'(i_x,j_x,k_y) \Tilde{\Delta} (j_x, k_y), \\
    K'(i_x,j_x,k_y) = \frac{1}{N_y} \sum_{k_x, k_x', k_y'} F(k_x,-k_y-k_y';k_x',k_y') \\
    \cross \omega_x ^{*} (i_x,k_x,-k_y-k_y')
    \omega_x^{*} (i_x,k_x',k_y')  \\
    \cross \omega_x (j_x,k_x,-k_y-k_y')
    \omega_x (j_x,k_x',k_y').
\end{gathered}
\end{equation}
The matrix $K'$ should be diagonalized for each value $k_y$, and the largest eigenvalue among them corresponds to $V^{-1}$.
The procedure has complexity $\mathcal{O} \left( N_x^4 N_y^2\right)$, nevertheless it could be reduced to $\mathcal{O} \left( N_x^4 N_y\right)$ assuming the fact that the order parameter should be constant in the $y$ direction.
It allows one to compute eigenvalues of $K'$ only for $k_y = 0$ (or $2\pi$).
Generalization of the three dimensional case (with TB parallel to the $yz$ plane) is straightforward. The calculation has complexity $\mathcal{O} \left( N_x^4 N_y N_z\right)$.
However, if one considers spin imbalance, the described approach ($k_y = 0$ assumption) cannot be used because the system can have non-uniform pair-density-wave as a ground state in some parameter range \cite{samoilenka2020pair,samoilenka2021microscopic,fulde1964superconductivity,larkin1965zh}.

\subsection{Only nearest neighbor hoppings}
If one considers electron hopping only between nearest neighbors in \eqref{eq:mean-field_Hamiltonian}, there is no critical temperature change due to TB presence (if it does not change coordination number).
The result also comes from the fact that the gap distribution should be uniform in the TB plane.
Consequently, Bogoliubov quasiparticles (wave functions) with only zero $k_y$ momentum contribute.
Therefore, all unit cells have identical phase at fixed position $x$, and kinetic terms now depend only on the number of links and their type.
If a TB does not change the lattice parameters, the Hamiltonian in the bulk and on a TB is the same; hence, $T_c$ remains the same.
Various numerical calculations also support the conclusion.
A specific example of the Hamiltonian calculation for the bulk and TB for a two dimensional rectangular lattice is given in \appref{app:Hamiltonian 2D}.

The result holds for homogeneous interaction strength and chemical potential.
In more general setting if one varies parameters of the model on a TB, it changes superconducting and transport properties.
For example, in Ref. \cite{zhitomirsky1997electronic}, authors concluded that in a normal state quasiparticles in considered here lattice (\figref{fig:2D rectangular lattice}) scatter from TB only for chemical potential on TB different from the bulk value.
It means that the electronic spectrum (wave functions and eigenenergies) does not change for uniform $\mu$ in the system, which is consistent with our conclusions.

\begin{figure*}
    \centering
    \includegraphics[width=0.8\linewidth]{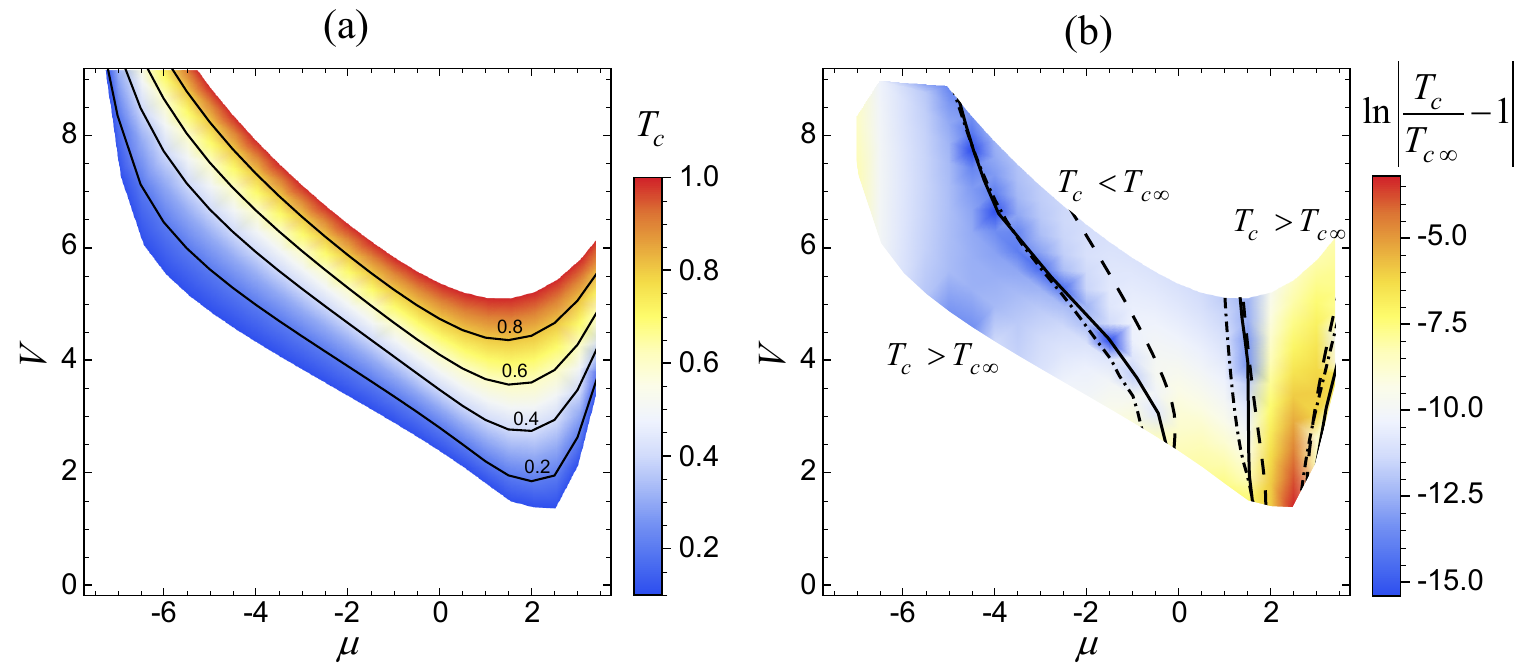}
    \caption{(a) Superconductivity phase diagram for a rectangular lattice with two TBs and periodic boundary conditions (\figref{fig:2D lattice model with NNNH and TB}), where $\mu$ is the chemical potential, $V$ is the onsite attraction potential.
    Solid lines correspond to constant critical temperature curves with $T_c$ written next to the line.
    Nearest neighbor hopping parameters are $t^{(1)}_1 = 1, \, t^{(1)}_2 = 1.4$, next-to-nearest hopping is $t^{(2)} = 7/12 \approx 0.58$, and the hopping across TB is $t^{(2)}_\text{TB} = 0.6$.
    \newline
    (b) Absolute value of relative change in the critical temperature for the system with two TBs ($T_c$) in comparison to the TB-free infinite system ($T_{c \infty}$).
    Solid lines are the “transition lines” between the regions where TBs lead to enhancement or suppression of $T_c$.
    By $T_c$ suppression we mean that global critical temperature of a system with parallel TBs is lower than $T_c$ in the TB-free case.
    Dashed and dot-dashed lines are the “transition lines” estimation obtained by LDOS argument and critical temperature comparison of TB-free infinite system and TB-like system (\figref{fig:2D TB-like lattice}), respectively.
    }
	\label{fig:2D phase diagram}
\end{figure*}

\section{Two dimensional rectangular lattice} \label{sec:Two dimensional rectangular lattice} \label{sec:2D Tc}

We consider a minimalistic model for a non-trivial cross-section of 110 TB in orthorhombic crystal (or 101 in tetragonal; for example, indium \cite{khlyustikov1985superconductivity}).
As discussed in a previous subsection, the critical temperature does not change due to the presence of TB for a rectangular lattice with nearest neighbors.
Therefore, we consider a lattice with nearest ($t^{(1)}_1, \, t^{(1)}_2$) and next-to-nearest neighbor [$t^{(2)} = t^{(1)}_1 t^{(1)}_2/(t^{(1)}_1 + t^{(1)}_2)$] hoppings (\figref{fig:2D lattice model with NNNH and TB}).
Where we used $t \propto r^{-2}$ hopping strength dependence.
The next-to-nearest inter-TB hopping (shown in orange in \figref{fig:2D lattice model with NNNH and TB}) has value $t^{(2)}_\text{TB} = (t^{(1)}_1 + t^{(1)}_2)/4$ in the model.
It is easy to show that $t^{(2)}_\text{TB} > t^{(2)}$ for $t^{(1)}_1 \neq t^{(1)}_2$.
For simplicity, we will use $t^{(1)}_1 = 1$ in the section, which can be interpreted as a new energy and temperature scale.
The system considered for numerical calculations has dimensions $400 \cross 200$ with periodic boundary conditions in both directions.
It models a system of parallel TBs (with a distance of 200 sites).

We apply the linearized gap equation approach [\eqref{eq:linearized gap equation final}] to investigate the superconducting phase diagram.
We consider $t^{(1)}_2 = 1.4$, which fixes all other hopping integrals, and we are left with three dependent parameters ($\mu, \, V, \, T_c$), which are illustrated on a phase diagram [\figref{fig:2D phase diagram}(a)].
The lower bound for chemical potential is $\mu=-7$ to restrict high interaction strength values (not to go too far beyond Bardeen-Cooper-Schrieffer limit), and the upper bound for $\mu$ is 3.4.
The chemical potential range covers all possible values of particle density in the system: From an empty band for $\mu = -7.13$ to a filled band for $\mu=3.13$ (in the case $T=0$).
Note the asymmetry of the diagram with respect to $\mu \rightarrow - \mu$ which indicates the absence of particle-hole symmetry in the system.
The asymmetry arises from the introduction of next-to-nearest neighbor hoppings.

Figure \ref{fig:2D phase diagram}(b) shows the relative change in the critical temperature compared to the infinite system (without TB, $T_{c \infty}$).
Notice that TBs lead to an increase in critical temperature for an almost empty (or filled) band, which is the opposite trend in contrast to the boundary critical temperature results \cite{samoilenka2020boundary, talkachov2023microscopic}.
One possible explanation is the following: For a TB, one hopping integral is increased. However, for a boundary, some of the hoppings are absent (which can be viewed as a significant amplitude decrease). 
Therefore, the overall trend on the phase diagrams should be reversed.
Solid black lines in \figref{fig:2D phase diagram}(b) separate regions with enhancement or suppression of $T_c$ due to the presence of TB.
Next, we present two approaches for the estimation of the transition lines (where TB presence does not change critical temperature, $T_c = T_{c \infty}$) on the $\left(\mu, \, V\right)$ diagram.

\begin{figure}[b]
    \begin{center}
		\includegraphics[width=0.99\columnwidth]{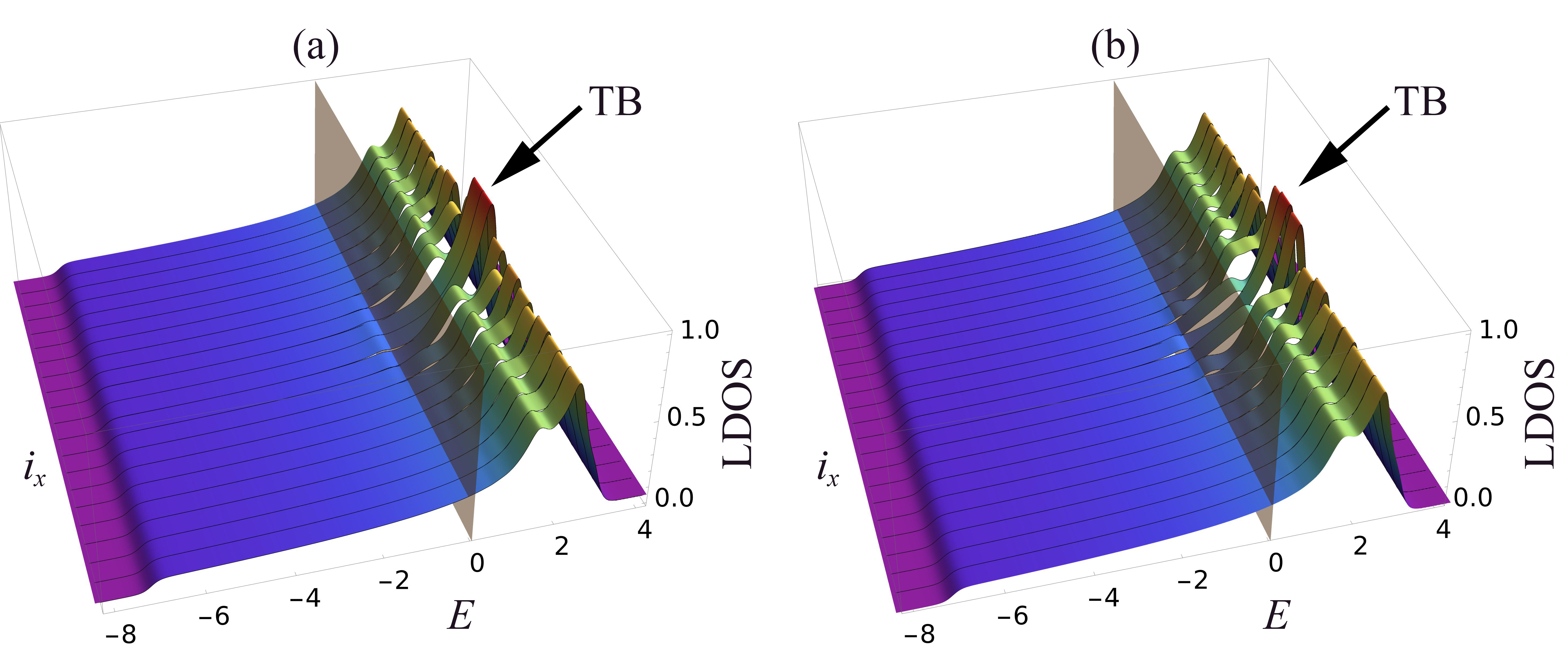}
		\caption{Local density of states oscillations in the vicinity of TB for a rectangular lattice in \figref{fig:2D lattice model with NNNH and TB}. Parameters of the system are $t^{(1)}_1=1, \, t^{(1)}_2=1.4, \, \mu = 0, \, T=0.1$ and no electron interaction ($V=0$). Panels (a) and (b) correspond to the sites along axes $x_1$ and $x_2$ [see \figref{fig:2D lattice model with NNNH and TB}(a)] respectively. The orange plane corresponds to the Fermi level.}
		\label{fig:2D LDOS}
\end{center}
\end{figure}

The first approach is related to the very short-range LDOS oscillations in the vicinity of TB.
The oscillations originate from deviations in electron wave functions from free electron case ($e^{i k x}$) due to TB presence.
The argument is based on computing DOS by averaging the LDOS oscillations in the macroscopically large volume \cite{samoilenka2020boundary, talkachov2023microscopic,SamoilenkaPhD}:
An increase in DOS (compared to bulk value) leads to critical temperature enhancement and vice versa.
The macroscopic region should include all the deviations in electron wave functions. However, the deviations are localized near TB (\figref{fig:2D LDOS}); therefore, the averaging region does not have to be large.
LDOS can be calculated for a TB in a normal state (\figref{fig:2D LDOS}).
We use averaging over the region of ten sites next to TB, which covers all the rapid oscillations (\figref{fig:2D LDOS}).
Using the algorithm, we calculated regions on a phase diagram [\figref{fig:2D phase diagram}(b)] where the critical temperature does not depend on the TB presence (shown with a dashed line).
Qualitative agreement with the linearized gap calculation results is observed, but the quantitative difference is noticeable.
It leads to the conclusion that LDOS is a crucial ingredient responsible for changes in critical temperature, though other factors have a significant influence.
It is essential to note that the LDOS oscillations emerge only in a full microscopic model.\
For instance, quasiclassic and Ginzburg-Landau approaches do not have the feature because of averaging all the short-length-scale physics for the model derivation. However, upon establishing superconductivity enhancement in a fully microscopic theory, one can microscopically derive additional interface terms for Ginzburg-Landau functional \cite{samoilenka2021microscopic}.

\begin{figure}[b]
    \begin{center}
		\includegraphics[width=0.35\columnwidth]{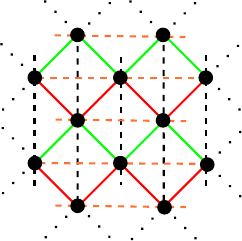}
		\caption{Infinite rectangular lattice that mimics TB region in \figref{fig:2D lattice model with NNNH and TB}.}
		\label{fig:2D TB-like lattice}
\end{center}
\end{figure}

The second possible explanation is based on the comparison of the superconducting critical temperature for the TB-free infinite system ($T_{c \infty}$) and TB-like system (which mimics most of the properties of TB, $T_{c \, \text{TB}}$).
The approach was introduced in the context of a different problem in Appendix D in Ref. \cite{talkachov2023microscopic}.
A superconducting state usually has lower energy than a normal state.
If we need to compare two superconducting states, a state with the highest $T_c$ is expected to have the lowest energy.
The analysis leads to the conclusion that TB leads to enhancement of critical temperature if $T_{c \, \text{TB}} > T_{c \infty}$ (for fixed chemical potential $\mu$ and interaction strength $V$) and vice versa.
The critical temperature for the TB-free infinite system ($T_{c \infty}$) can be found in a standard way using integration over the Brillouin zone.
Calculation of TB one ($T_{c \, \text{TB}}$), however, is more tricky because TB is effectively not a one dimensional object (like a chain of sites along a blue dashed line in \figref{fig:2D lattice model with NNNH and TB}) since the presentation does not correctly capture coordination number and, more importantly, does not include inter-TB hoppings.
One of the possible models of the TB-like structure is illustrated in \figref{fig:2D TB-like lattice}.
The twin boundary in \figref{fig:2D lattice model with NNNH and TB} is an element of such lattice.
We will use superconducting critical temperature for the lattice in \figref{fig:2D TB-like lattice} as the estimation for the TB critical temperature ($T_{c \, \text{TB}}$).
The computation of $T_{c \, \text{TB}}$ is presented in \appref{app:Hamiltonian TB-like lattice}.
The method gives transition lines (where critical temperatures for both systems are equal) illustrated with a dot-dashed line in \figref{fig:2D phase diagram}(b).
One can note much better quantitative agreement with linearized gap equation results.

To compare the theories in another parameter regime, we fix $\mu = 0$ and investigate different region of hopping $t_2^{(1)}$ and interaction strength $V$ parameter space (the figures are not presented in the paper).
In this regime, the linearized gap equation approach shows that TB always leads to critical temperature suppression ($T_c < T_{c \infty}$). 
Modulus of critical temperature deviation from the bulk value increases monotonically when increasing $t_2^{(1)}$ and/or decreasing $V$.
The approximate approach based on the comparison critical temperature for the TB-free infinite system ($T_{c \infty}$) and TB-like system ($T_{c \, \text{TB}}$) leads to qualitatively correct results which reproduce microscopic calculation of $T_c - T_{c \infty}$ behavior.
However, the approximate LDOS argument does not give the desired result: The difference in LDOS has non-monotonic behavior when increasing temperature and, moreover, there is a narrow parameter range $t_2^{(1)} \in (1; 1.2), \, T_c \in (0.2; 0.25)$ where the approach predicts $T_c > T_{c \infty}$.

In the section, we considered a fixed-size periodic system with two TBs. However, two   questions arise:
(i) How does the critical temperature deviation from the TB-free case depend on the distance between TBs?
(ii) When we studied two TBs, we omitted next-to-nearest TB interactions. Do the interactions give non-negligible contributions?
We address the questions in \appref{app:Tc dependence on the period}.

\section{Body centered cubic lattice} \label{sec:Body centered cubic lattice}

\begin{figure}
    \centering
    \includegraphics[scale=0.3]{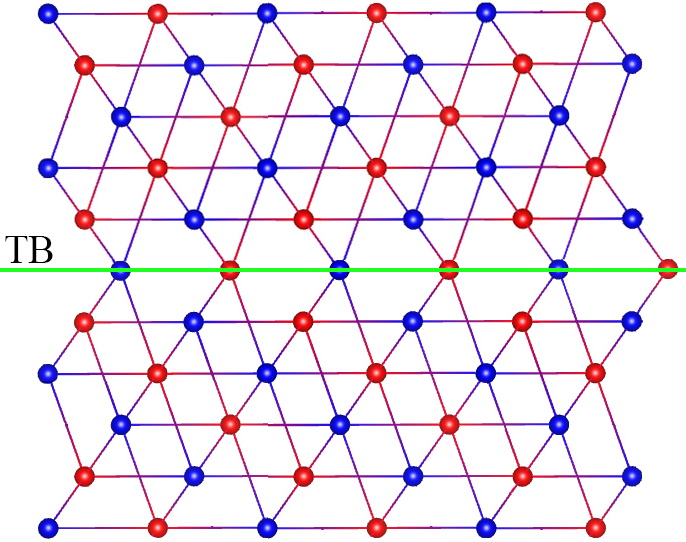}
    \caption{A 112 body centered cubic crystal TB. The two sublattices of the BCC crystal are shown in red and blue.}
    \label{fig:bcctwin}
\end{figure}

The section deals with 112 TB in the BCC crystal (the example is Nb \cite{khlyustikov1985superconductivity}), as depicted in \figref{fig:bcctwin}.
For the TB type, hopping integrals obey changes similar to the rectangular lattice considered earlier: An atom lying on a TB has the same coordination number and distances to neighbors as a bulk atom; however, an atom one layer away has 5 (out of 6) regular nearest neighbors, 7 (out of 8) regular next-to-nearest neighbors, and a link across TB (which is shorter than nearest neighbor distance).
We introduce a sharp cutoff for hopping integrals, corresponding to distances greater than next-to-nearest neighbor separation.
Using the hopping notation introduced earlier and assuming $t \propto r^{-2}$ dependence, we can write $t^{(1)} = 4/3, \, t^{(1)}_\text{TB} = 1.5$ using $t^{(2)} = 1$ as a new energy unit.
An infinite crystal with the parameters has an energy band $[-16.67; \, 6]$.
The typical system size considered is $400 \cross 100 \cross 100$ with periodic boundary conditions [similar to \figref{fig:2D lattice model with NNNH and TB}(a), but now TB forms the $yz$ plane].

\begin{figure*}
    \begin{center}
		\includegraphics[width=0.8\linewidth]{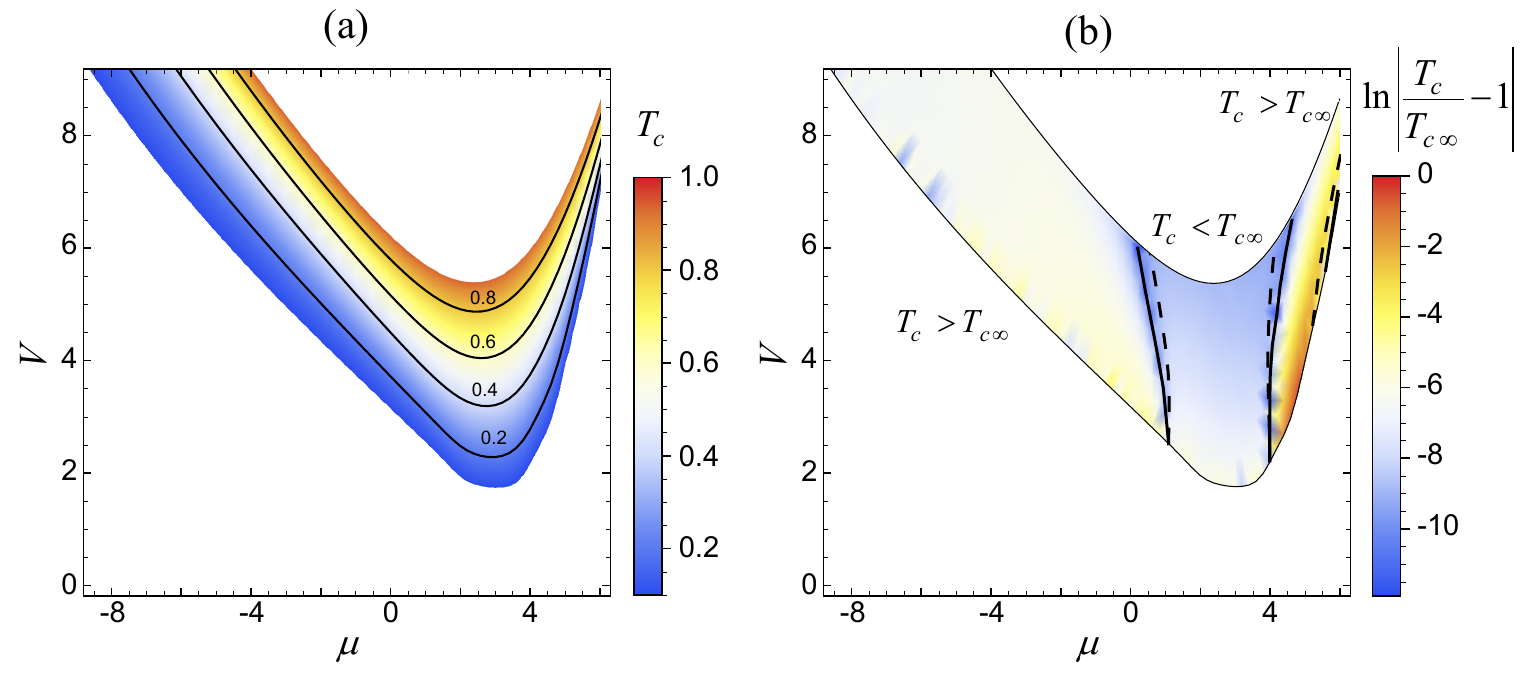}
    \caption{(a) Superconductivity phase diagram for BCC crystal with 112 TBs (\figref{fig:bcctwin}), where $\mu$ is the chemical potential, $V$ is the onsite attraction potential.
    Solid lines correspond to constant critical temperature curves with $T_c$ written next to the line.
    Nearest neighbor and next-to-nearest hopping integrals are $t^{(1)} = 4/3$ and $t^{(2)} = 1$ respectively; the hopping across TB is $t^{(1)}_\text{TB} = 1.5$.
    \newline
    (b) Absolute value of relative change in the critical temperature for the system with two TBs ($T_c$) in comparison to the TB-free infinite lattice ($T_{c \infty}$).
    Thin solid lines show $\mu, \, V$ region of the calculations.
    Bold solid lines are the “transition lines” between the regions where TBs lead to enhancement or suppression of $T_c$.
    Dashed lines are the estimation of the “transition lines” using the LDOS argument.}
     \label{fig:BCC phase diagram}
	\end{center}
\end{figure*}

We study the superconducting phase diagram obtained using the linearized gap equation approach [\eqref{eq:linearized gap equation final}].
Results are illustrated in \figref{fig:BCC phase diagram}.
Again, we restricted the plotted region to have a  relatively low interaction strength ($V<9$), which made the low band filling region inaccessible ($\mu < -8.5$, which corresponds to less than 0.15 electrons per site).
There is a critical temperature suppression for an almost empty band ($T_c < T_{c \infty}$ for $\mu \lesssim -12$), which is out of range in \figref{fig:BCC phase diagram}(b).
The overall trend in \figref{fig:BCC phase diagram}(b) is somewhat similar as in the superconductor-insulator boundary problems \cite{samoilenka2020boundary,talkachov2023microscopic}: Large intermediate doping regime with $T_c > T_{c \infty}$ and critical temperature change decrease close to filled/empty band.
However, it has an exception: The region $\mu \in [1;4]$ where $T_c < T_{c \infty}$ which qualitatively mimics results from the previous section [\figref{fig:2D phase diagram}(b)].

To understand the origin of the exception (is it coming from the link $t^{(1)}_\text{TB}$ or related to the absence of next-to-nearest neighbor), we calculate the system with $t^{(1)}_\text{TB} = t^{(1)}$ which captures consequences only of the second effect.
The critical temperature change phase diagram is quantitatively similar to \figref{fig:BCC phase diagram}(b), which indicates that enhanced hopping across TB gives a small contribution.
It has an explanation: A link across TB that substitutes for one of the nearest neighbors has a relatively weak effect compared to the absence of the next-to-nearest neighbor.
In the first case, we could say that one of the hopping integrals effectively changes from $t^{(1)} = 4/3$ to $t^{(1)}_\text{TB} = 1.5$.
Nevertheless, in the second case, we go from $t^{(2)} = 1$ to zero value when TB is present.
This change is roughly one order of magnitude greater than in the first case; therefore, the second effect is dominant in most of the phase diagram.

We checked the applicability of the LDOS argument in the case [dashed lines in \figref{fig:BCC phase diagram}(b)] by analogy with the previous section.
One could note good quantitative agreement with the calculation results.
Notice quite dramatic enhancement in critical temperature compared to the infinite system in the vicinity $\mu \approx 5$, which also corresponds to a spike on the LDOS diagram (van Hove singularity).
In the region, there is a clear correlation between a change in LDOS and a change in critical temperature (the figure is not presented in the paper).
The rapid increase in LDOS is the consequence of the band touching the Fermi surface at the point H ($k_x=0, \, k_y=2\pi, \, k_z = 0$).

An alternative method (proposed in \secref{sec:2D Tc}) for the estimation of ``transition lines" on a phase diagram does not apply to the three dimensional lattice.
The reason is that there are two different atom types (with different coordination numbers) for a three dimensional TB-like lattice in the case.
It leads to non-uniform gap distribution even for infinite system which forbids to use equation like \eqref{eq:V calculation}.

The general trend on the phase diagram [\figref{fig:BCC phase diagram}(b)] is similar to behavior of phase diagrams for boundary superconducting critical temperature change \cite{talkachov2023microscopic,samoilenka2020boundary}: critical temperature enhancement for moderate band filling and suppression for an almost empty/filled band. The similarities originate from the absence of one of the neighbors. The only drastic exception (difference) for the BCC lattice comes from a region with van Hove singularity where one could obtain significant critical temperature increase.

\section{Conclusion}
The experimentally observed change in superconducting critical temperature on twin boundaries is often conjectured to be related to the possible change in the interaction on TB arising from modified phonons \cite{nabutovskij1981localized,averin1983theory}.
In the paper, we have shown that modification of critical temperature on twin boundaries can arise   from ``\textit{geometric effects}",  connected to properties of electron scattering and interference on a defect. We show that geometric effects can both enhance or suppress $T_c$ depending on the chemical potential and electron interaction strength.

It was shown that $T_c$ in the presence of TB should not change for the considered models with nearest neighbor hoppings if there are no TB modifications and no coordination number change.
For the two dimensional rectangular lattice (mimicking 110 twin boundary in orthorhombic crystal) considered with next-to-nearest neighbor hoppings, we obtained a superconductivity critical temperature change phase diagram [\figref{fig:2D phase diagram}(b)], which shows suppression of $T_c$ (in comparison to infinite TB-free lattice) for moderate electron filling and enhancement close to an empty/filled band.
Applying the argument of short-length-scale LDOS oscillations close to the defect  \cite{talkachov2023microscopic,samoilenka2020boundary,SamoilenkaPhD} gives only qualitative correct results for the question of critical temperature change due to TB.
To address the question, we developed another method based on critical temperature comparison for the bulk and TB-like structure (\figref{fig:2D TB-like lattice}).
The method gives better quantitative agreement with calculation results [\figref{fig:2D phase diagram}(b)].
Comparing the phase diagram [\figref{fig:2D phase diagram}(b)] to phase diagrams for boundary superconducting critical temperature change \cite{talkachov2023microscopic,samoilenka2020boundary}, one could note qualitatively opposite behavior.
The reason for this is the hopping integral across TB enhancement (compared to the bulk value for the link), unlike the boundaries where some hoppings are decreased to zero.

The three dimensional model for body centered cubic crystal with 112 TB has an interplay of two effects: An increase in the hopping across the TB and the absence of the next-to-nearest neighbor.
The second reason gives a dominant contribution to the deviation of superconducting critical temperature from the bulk one.
The phase diagram has a large chemical potential and electron interaction strength parameter space region with critical temperature enhancement.
There is a large critical temperature enhancement for a chemical potential $\mu \approx 5$, related to the high density of states due to the band touching the Fermi surface.
There is a high correlation between averaged very short-range LDOS oscillations in the vicinity of TB and critical temperature deviation from TB-free case for the BCC crystal.

We can highlight the main ingredients for the geometric-factors-based critical temperature change due to the presence of defects.
Coordination number reduction for nearest neighbor hoppings leads to critical temperature decrease close to filled/empty band and increase for moderate band filling  \cite{talkachov2023microscopic,samoilenka2020boundary,SamoilenkaPhD}.
Increasing one of the hopping integrals (for example, due to TB) leads to qualitatively opposite results for high temperatures; however, for a low temperatures phase diagram (\figref{fig:2D phase diagram}) is more diverse.
Coordination number reduction for next-to-nearest neighbor hoppings also leads to a more diverse phase diagram (\figref{fig:BCC phase diagram}) with a region of critical temperature suppression for intermediate filling.
In the considered cases the interplay of both effects can be qualitatively predicted using argument of averaging of geometrically-induced variation of short-scale LDOS oscillations next to the defect.
The works show the importance of geometric effects and short-scale physics for enhancement or suppression of superconductivity at a macroscopic length scale away from a twinning plane.
It also suggests that short-length scale physics and geometric effects  can also be significant for other types of defects.

\begin{acknowledgments}
We thank Albert Samoilenka and Mats Barkman for useful discussions.
This work was supported by the Knut and Alice Wallenberg Foundation via the Wallenberg Center for Quantum Technology (WACQT) and Swedish Research Council Grant
2022-04763,  
Carl Trygger Foundation through the grant CTS 20:25, Olle Engkvists
Stiftelse.  
Nordita is supported in part by NordForsk.

\end{acknowledgments}

\appendix

\section{Hamiltonian calculation for the bulk and TB for two dimensional rectangular lattice} \label{app:Hamiltonian 2D}

\begin{figure}
    \begin{center}
		\includegraphics[width=0.8\columnwidth]{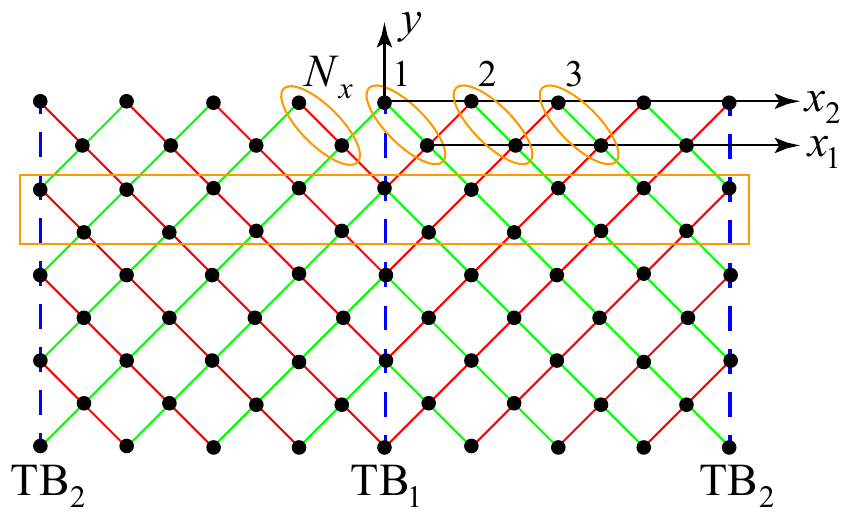}
		\caption{Periodic in both directions rectangular lattice with two TBs, which models a system with an infinite series of TBs. TBs are indicated with blue dashed lines. Orange ovals show elementary unit cells. The orange rectangle shows the supercell for Fourier transform in the $y$ direction and numerical calculations.}
		\label{fig:2D rectangular lattice}
\end{center}
\end{figure}

The kinetic part (assuming $\mu = 0$) of Hamiltonian \eqref{eq:mean-field_Hamiltonian} applied to the infinite rectangular lattice with nearest neighbor hoppings (right part of the system in \figref{fig:2D rectangular lattice}) reads

\begin{equation}
\begin{gathered}
    H = -\sum_{i_x, i_y \notin \text{TB}} \medmath{c_2 \da (i_x,i_y) \left( t_\text{green}\left[ c_1 (i_x,i_y) + c_1 (i_x-1,i_y+1) \right] \right.} \\
    + \left. t_\text{red}\left[c_1 (i_x-1,i_y) + c_1 (i_x,i_y+1)  \right] \right)+\text{h.c.},
\end{gathered}
\end{equation}
where $c_{\alpha}\da (c_{\alpha})$ is the electron creation (annihilation) operator on axis $x_\alpha$ (see \figref{fig:2D rectangular lattice}).

If one goes to momentum space only in the $y$ direction by performing $c_\alpha (i_x,i_y) = \frac{1}{\sqrt{N_y/2}} \sum_{k_y} e^{- \ii k_y i_y} \Tilde{c}_\alpha (i_x,k_y)$ the Hamiltonian transforms to
\begin{equation} \label{eq:2D Hamiltonian bulk}
\begin{gathered}
    H = \medmath{-\sum_{i_x\notin \text{TB}, k_y} \Tilde{c}_2 \da (i_x,k_y) \left( t_\text{green}\left[ \Tilde{c}_1 (i_x,k_y) + \Tilde{c}_1 (i_x-1,k_y) e^{-\ii k_y} \right] \right.} \\
    + \left. t_\text{red}\left[\Tilde{c}_1 (i_x-1,k_y) + \Tilde{c}_1 (i_x,k_y) e^{-\ii k_y}  \right] \right)+\text{h.c.}
\end{gathered}
\end{equation}
The result for the left part of the system illustrated in \figref{fig:2D rectangular lattice} is obtained by swap $t_\text{green} \leftrightarrow t_\text{red}$.

If one writes part of the Hamiltonian for sites on TB (they belong to the axis $x_2$):
\begin{equation}
\begin{gathered}
    H_\text{TB} = \medmath{-\sum_{i_x, i_y \in \text{TB}} c_2 \da (i_x,i_y) \left( t_\text{green}\left[ c_1 (i_x,i_y) + c_1 (i_x-1,i_y)  \right] \right.} \\
    + \left. t_\text{red}\left[c_1 (i_x,i_y+1) + c_1 (i_x-1,i_y+1)  \right] \right)+\text{h.c.}
\end{gathered}
\end{equation}
and in momentum space
\begin{equation} \label{eq:2D Hamiltonian TB}
\begin{gathered}
    H_\text{TB}  = \medmath{-\sum_{i_x \in \text{TB}, k_y}  \Tilde{c}_2 \da (i_x,k_y) \left( t_\text{green}\left[ \Tilde{c}_1 (i_x,k_y) + \Tilde{c}_1 (i_x-1,k_y) \right] \right.} \\
    + \left. t_\text{red} e^{-\ii k_y} \left[\Tilde{c}_1 (i_x,k_y) + \Tilde{c}_1 (i_x-1,k_y)   \right] \right)+\text{h.c.}
\end{gathered}
\end{equation}

Expressions \eqref{eq:2D Hamiltonian bulk} and \eqref{eq:2D Hamiltonian TB} coincide for $k_y = 0$ (or $2 \pi)$ which is the only relevant momentum contribution for the considered superconductivity model (see \secref{sec:Superconductivity model}).
Therefore, Hamiltonian is homogeneous in the system, which means that rectangular lattice with nearest neighbors hoppings does not ``feel" TB presence.

\section{Hamiltonian calculation for two dimensional TB-like infinite lattice} \label{app:Hamiltonian TB-like lattice}

For a lattice illustrated in \figref{fig:2D TB-like lattice}, one can write Hamiltonian in the following form (using site indexing  similarly to \figref{fig:2D rectangular lattice}, $t_\text{green} = t^{(1)}_1 , \, t_\text{red} = t^{(1)}_2$)
\begin{equation}
\begin{gathered}
    H = - \sum_{i_x, i_y} \left( c_1 \da (i_x,i_y) \left( t_\text{green} \left[ c_2 (i_x,i_y) + c_2 (i_x+1,i_y)\right] \right.\right.\\
    \left. + t_\text{red} \left[ c_2 (i_x,i_y-1) + c_2 (i_x+1,i_y-1)\right] \right)  + \text{h.c.}\\
    + \sum_{\alpha=1}^2 c_\alpha \da (i_x,i_y) \left(
    t_\text{TB}^{(2)} \left[c_\alpha (i_x-1,i_y) + c_\alpha (i_x+1,i_y)\right] \right. 
    \\
    \left. \left. + t^{(2)} \left[c_\alpha (i_x,i_y-1) + c_\alpha (i_x,i_y+1)\right]  + \mu c_\alpha (i_x,i_y)\right)
    \right).
\end{gathered}
\end{equation}

The Hamiltonian in momentum space $c_\alpha (i_x,i_y) = \frac{1}{\sqrt{N_x N_y}} \sum_{k_x,k_y} e^{- \ii (k_x i_x + k_y i_y)} \Tilde{c}_\alpha (k_x,k_y)$ reads as
\begin{widetext}
    \begin{equation}
        H = - \sum_{k_x, k_y} \left(
        \begin{matrix}
            \Tilde{c}_1 \da (k_x,k_y) & \Tilde{c}_2 \da (k_x,k_y)
        \end{matrix} \right) 
        \left(
        \begin{matrix}
            \mu + 2 \left( t_\text{TB}^{(2)} \cos k_x + t^{(2)} \cos k_y\right) & \left( 1 + e^{- \ii k_x}\right) \left(t_\text{green} + t_\text{red} e^{\ii k_y} \right) \\
            \left( 1 + e^{ \ii k_x}\right) \left(t_\text{green} + t_\text{red} e^{-\ii k_y} \right) & \mu + 2 \left( t_\text{TB}^{(2)} \cos k_x + t^{(2)} \cos k_y\right)
        \end{matrix} \right)
        \left(
        \begin{matrix}
            \Tilde{c}_1 (k_x,k_y) \\
            \Tilde{c}_2 (k_x,k_y)
        \end{matrix} \right).
    \end{equation}
\end{widetext}
It has the following eigenvalues:
\begin{equation}
\begin{gathered}
    E_s = -\mu - 2 \left( t_\text{TB}^{(2)} \cos k_x + t^{(2)} \cos k_y\right) \\
    - s \sqrt{\left(t_\text{green}^2 + t_\text{red}^2 + 2 t_\text{green} t_\text{red} \cos k_y \right) \cos^2 \left(\frac{k_x}{2}\right)}
\end{gathered}
\end{equation}
where $s=\pm 1$. To compute critical temperature, one can use the integral equation (for given $\mu$, $V$ and assuming $\Delta = 0$ at the phase transition)
\begin{equation} \label{eq:V calculation}
    V^{-1} = \frac{1}{2 S_\text{1st BZ}} \int_\text{1st BZ} dk_x dk_y \sum_{s = \pm 1} \frac{\tanh{\frac{E_s}{2 T}}}{E_s} .
\end{equation}
The calculation is convenient to perform after applying transformation $k_x \rightarrow k_x'+k_y'$, $k_y \rightarrow k_x'-k_y'$ which rotates the system by $\pi/4$ and makes 1st BZ a square with $2\pi$ sides which are oriented along $k_x'$ and $k_y'$ respectively.

\section{Critical temperature change dependence on twin boundary spacing}  \label{app:Tc dependence on the period}

The study of the relative change in the critical temperature on the TB spacing $L_\text{TBs}$ allows one to understand TB interactions better.
Namely, we used TB spacing equal to 200 layers in the main text. Was it sufficient spacing to omit next-to-nearest TB interactions? We bring light to the problems in the appendix.

In order to get a numerical answer to the question, we calculate a two dimensional lattice in two ways:
\begin{itemize}
    \item Varying the size of the system [\figref{fig:2D lattice model with NNNH and TB}(a)] in the $x$ direction;
    \item Fixing $N_x=1200$ and varying the number of TBs in the system (keeping it even; TBs are equidistant).
\end{itemize}

\begin{figure}[b]
    \begin{center}
		\includegraphics[width=0.99\columnwidth]{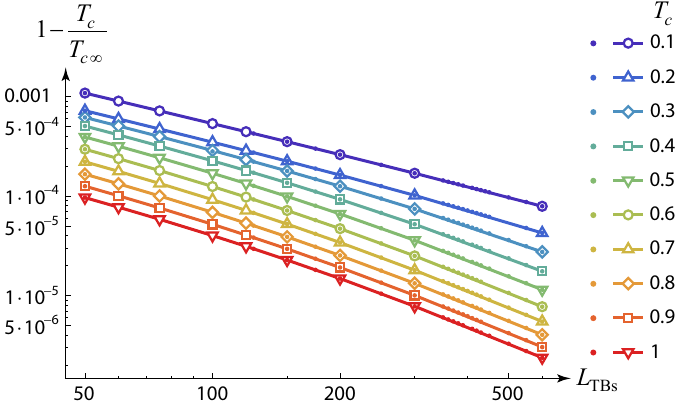}
		\caption{The dependence of relative change in critical temperature (in comparison to infinite TB-free system) for rectangular lattice on the distance between TBs ($L_\text{TBs}$). 
                Dots correspond to the system of two TBs with periodic boundary conditions [\figref{fig:2D lattice model with NNNH and TB}(a)], line with open figures corresponds to a system with $N=1200/L_\text{TBs}$ TBs.
                Parameters are $t^{(1)}_1=1, \, t^{(1)}_2=1.4, \, \mu = 0$, which correspond to critical temperature suppression due to TB presence.}
		\label{fig:2D dTc(L)}
\end{center}
\end{figure}

Figure \ref{fig:2D dTc(L)} shows the dependence $1 - T_c/T_{c \infty}$ on the TB spacing obtained by both methods.
For the two methods, results coincide up to the computational error.
The exception is the point $T_c=0.1, \, L_\text{TBs} = 50$, which means that the effect of next-to-nearest TB interaction is negligible for $L_\text{TBs} \geq 100$.
We observe the interaction for the lowest considered temperature because LDOS oscillations near TB (\figref{fig:2D LDOS}) cover larger space region when $T_c$ decreases.
Note that in the log-log scale, points lie on a straight line only for the lowest considered temperature ($T_c=0.1$).
The higher the critical temperature, the larger bending in \figref{fig:2D dTc(L)}, which means non-power law behavior.
For the lowest temperature considered ($T_c=0.1$) power $\alpha$ is close to the $-1$.
It is consistent with results of Ginzburg-Landau theory \cite{averin1983theory} and approximate microscopic calculations \cite{suslov1989mechanism} in a certain parameter regime.

The main point of the appendix:  Next-to-nearest TB interactions give notable contribution for the lowest considered temperature ($T_c = 0.1$) with TB spacing of fifty sites. The distance is much smaller than what we chose for calculations in the main text. Therefore, the main results of the paper are reliable.

\bibliography{references}

\begin{thebibliography}{35}%
\makeatletter
\providecommand \@ifxundefined [1]{%
 \@ifx{#1\undefined}
}%
\providecommand \@ifnum [1]{%
 \ifnum #1\expandafter \@firstoftwo
 \else \expandafter \@secondoftwo
 \fi
}%
\providecommand \@ifx [1]{%
 \ifx #1\expandafter \@firstoftwo
 \else \expandafter \@secondoftwo
 \fi
}%
\providecommand \natexlab [1]{#1}%
\providecommand \enquote  [1]{``#1''}%
\providecommand \bibnamefont  [1]{#1}%
\providecommand \bibfnamefont [1]{#1}%
\providecommand \citenamefont [1]{#1}%
\providecommand \href@noop [0]{\@secondoftwo}%
\providecommand \href [0]{\begingroup \@sanitize@url \@href}%
\providecommand \@href[1]{\@@startlink{#1}\@@href}%
\providecommand \@@href[1]{\endgroup#1\@@endlink}%
\providecommand \@sanitize@url [0]{\catcode `\\12\catcode `\$12\catcode `\&12\catcode `\#12\catcode `\^12\catcode `\_12\catcode `\%12\relax}%
\providecommand \@@startlink[1]{}%
\providecommand \@@endlink[0]{}%
\providecommand \url  [0]{\begingroup\@sanitize@url \@url }%
\providecommand \@url [1]{\endgroup\@href {#1}{\urlprefix }}%
\providecommand \urlprefix  [0]{URL }%
\providecommand \Eprint [0]{\href }%
\providecommand \doibase [0]{https://doi.org/}%
\providecommand \selectlanguage [0]{\@gobble}%
\providecommand \bibinfo  [0]{\@secondoftwo}%
\providecommand \bibfield  [0]{\@secondoftwo}%
\providecommand \translation [1]{[#1]}%
\providecommand \BibitemOpen [0]{}%
\providecommand \bibitemStop [0]{}%
\providecommand \bibitemNoStop [0]{.\EOS\space}%
\providecommand \EOS [0]{\spacefactor3000\relax}%
\providecommand \BibitemShut  [1]{\csname bibitem#1\endcsname}%
\let\auto@bib@innerbib\@empty
\bibitem [{\citenamefont {Nabutovskij}\ and\ \citenamefont {Shapiro}(1981)}]{nabutovskij1981localized}%
  \BibitemOpen
  \bibfield  {author} {\bibinfo {author} {\bibfnamefont {V.}~\bibnamefont {Nabutovskij}}\ and\ \bibinfo {author} {\bibfnamefont {B.~Y.}\ \bibnamefont {Shapiro}},\ }\bibfield  {title} {\bibinfo {title} {Localized superconducting states in inhomogeneous superconductors},\ }\href@noop {} {\bibfield  {journal} {\bibinfo  {journal} {Fiz. Nizk. Temp.(Kiev);(Ukrainian SSR)}\ }\textbf {\bibinfo {volume} {7}} (\bibinfo {year} {1981})}\BibitemShut {NoStop}%
\bibitem [{\citenamefont {Averin}\ \emph {et~al.}(1983)\citenamefont {Averin}, \citenamefont {Buzdin},\ and\ \citenamefont {Bulaevskii}}]{averin1983theory}%
  \BibitemOpen
  \bibfield  {author} {\bibinfo {author} {\bibfnamefont {V.}~\bibnamefont {Averin}}, \bibinfo {author} {\bibfnamefont {A.}~\bibnamefont {Buzdin}},\ and\ \bibinfo {author} {\bibfnamefont {L.}~\bibnamefont {Bulaevskii}},\ }\bibfield  {title} {\bibinfo {title} {Theory of localized superconductivity},\ }\href@noop {} {\bibfield  {journal} {\bibinfo  {journal} {Zh. Eksp. Teor. Fiz}\ }\textbf {\bibinfo {volume} {84}},\ \bibinfo {pages} {748} (\bibinfo {year} {1983})}\BibitemShut {NoStop}%
\bibitem [{\citenamefont {{Andreev}}(1987)}]{andreev1987Exotic}%
  \BibitemOpen
  \bibfield  {author} {\bibinfo {author} {\bibfnamefont {A.~F.}\ \bibnamefont {{Andreev}}},\ }\bibfield  {title} {\bibinfo {title} {{Exotic superconductivity of twinning planes}},\ }\href@noop {} {\bibfield  {journal} {\bibinfo  {journal} {Pisma v Zhurnal Eksperimentalnoi i Teoreticheskoi Fiziki}\ }\textbf {\bibinfo {volume} {46}},\ \bibinfo {pages} {463} (\bibinfo {year} {1987})}\BibitemShut {NoStop}%
\bibitem [{\citenamefont {Abrikosov}\ and\ \citenamefont {Buzdin}(1988)}]{abrikosov1988manifestation}%
  \BibitemOpen
  \bibfield  {author} {\bibinfo {author} {\bibfnamefont {A.}~\bibnamefont {Abrikosov}}\ and\ \bibinfo {author} {\bibfnamefont {A.~I.}\ \bibnamefont {Buzdin}},\ }\bibfield  {title} {\bibinfo {title} {Manifestation of superconductivity of the twinning planes of high-temperature superconductors},\ }\href@noop {} {\bibfield  {journal} {\bibinfo  {journal} {JETP LETTERS}\ }\textbf {\bibinfo {volume} {47}},\ \bibinfo {pages} {247} (\bibinfo {year} {1988})}\BibitemShut {NoStop}%
\bibitem [{\citenamefont {Dimos}\ \emph {et~al.}(1990)\citenamefont {Dimos}, \citenamefont {Chaudhari},\ and\ \citenamefont {Mannhart}}]{dimos1990superconducting}%
  \BibitemOpen
  \bibfield  {author} {\bibinfo {author} {\bibfnamefont {D.}~\bibnamefont {Dimos}}, \bibinfo {author} {\bibfnamefont {P.}~\bibnamefont {Chaudhari}},\ and\ \bibinfo {author} {\bibfnamefont {J.}~\bibnamefont {Mannhart}},\ }\bibfield  {title} {\bibinfo {title} {Superconducting transport properties of grain boundaries in yba 2 cu 3 o 7 bicrystals},\ }\href@noop {} {\bibfield  {journal} {\bibinfo  {journal} {Physical Review B}\ }\textbf {\bibinfo {volume} {41}},\ \bibinfo {pages} {4038} (\bibinfo {year} {1990})}\BibitemShut {NoStop}%
\bibitem [{\citenamefont {Bobrov}\ and\ \citenamefont {Lebyodkin}(1991)}]{BOBROV1991411}%
  \BibitemOpen
  \bibfield  {author} {\bibinfo {author} {\bibfnamefont {V.}~\bibnamefont {Bobrov}}\ and\ \bibinfo {author} {\bibfnamefont {M.}~\bibnamefont {Lebyodkin}},\ }\bibfield  {title} {\bibinfo {title} {Possible twin-boundary effect upon temperature dependence of critical current near tc of high-tc superconductors},\ }\href {https://doi.org/https://doi.org/10.1016/0921-4534(91)90091-C} {\bibfield  {journal} {\bibinfo  {journal} {Physica C: Superconductivity}\ }\textbf {\bibinfo {volume} {178}},\ \bibinfo {pages} {411} (\bibinfo {year} {1991})}\BibitemShut {NoStop}%
\bibitem [{\citenamefont {Nabatame}\ \emph {et~al.}(1994)\citenamefont {Nabatame}, \citenamefont {Koike}, \citenamefont {Hyun}, \citenamefont {Hirabayashi}, \citenamefont {Suhara},\ and\ \citenamefont {Nakamura}}]{nabatame1994transport}%
  \BibitemOpen
  \bibfield  {author} {\bibinfo {author} {\bibfnamefont {T.}~\bibnamefont {Nabatame}}, \bibinfo {author} {\bibfnamefont {S.}~\bibnamefont {Koike}}, \bibinfo {author} {\bibfnamefont {O.}~\bibnamefont {Hyun}}, \bibinfo {author} {\bibfnamefont {I.}~\bibnamefont {Hirabayashi}}, \bibinfo {author} {\bibfnamefont {H.}~\bibnamefont {Suhara}},\ and\ \bibinfo {author} {\bibfnamefont {K.}~\bibnamefont {Nakamura}},\ }\bibfield  {title} {\bibinfo {title} {Transport superconducting properties of grain boundaries in tl1ba2ca2cu3o x thin films},\ }\href@noop {} {\bibfield  {journal} {\bibinfo  {journal} {Applied physics letters}\ }\textbf {\bibinfo {volume} {65}},\ \bibinfo {pages} {776} (\bibinfo {year} {1994})}\BibitemShut {NoStop}%
\bibitem [{\citenamefont {Song}\ \emph {et~al.}(2012{\natexlab{a}})\citenamefont {Song}, \citenamefont {Wang}, \citenamefont {Jiang}, \citenamefont {Wang}, \citenamefont {He}, \citenamefont {Chen}, \citenamefont {Hoffman}, \citenamefont {Ma},\ and\ \citenamefont {Xue}}]{PhysRevLett.109.137004}%
  \BibitemOpen
  \bibfield  {author} {\bibinfo {author} {\bibfnamefont {C.-L.}\ \bibnamefont {Song}}, \bibinfo {author} {\bibfnamefont {Y.-L.}\ \bibnamefont {Wang}}, \bibinfo {author} {\bibfnamefont {Y.-P.}\ \bibnamefont {Jiang}}, \bibinfo {author} {\bibfnamefont {L.}~\bibnamefont {Wang}}, \bibinfo {author} {\bibfnamefont {K.}~\bibnamefont {He}}, \bibinfo {author} {\bibfnamefont {X.}~\bibnamefont {Chen}}, \bibinfo {author} {\bibfnamefont {J.~E.}\ \bibnamefont {Hoffman}}, \bibinfo {author} {\bibfnamefont {X.-C.}\ \bibnamefont {Ma}},\ and\ \bibinfo {author} {\bibfnamefont {Q.-K.}\ \bibnamefont {Xue}},\ }\bibfield  {title} {\bibinfo {title} {Suppression of superconductivity by twin boundaries in fese},\ }\href {https://doi.org/10.1103/PhysRevLett.109.137004} {\bibfield  {journal} {\bibinfo  {journal} {Phys. Rev. Lett.}\ }\textbf {\bibinfo {volume} {109}},\ \bibinfo {pages} {137004} (\bibinfo {year} {2012}{\natexlab{a}})}\BibitemShut {NoStop}%
\bibitem [{\citenamefont {Watashige}\ \emph {et~al.}(2015)\citenamefont {Watashige}, \citenamefont {Tsutsumi}, \citenamefont {Hanaguri}, \citenamefont {Kohsaka}, \citenamefont {Kasahara}, \citenamefont {Furusaki}, \citenamefont {Sigrist}, \citenamefont {Meingast}, \citenamefont {Wolf}, \citenamefont {L\"ohneysen}, \citenamefont {Shibauchi},\ and\ \citenamefont {Matsuda}}]{PhysRevX.5.031022}%
  \BibitemOpen
  \bibfield  {author} {\bibinfo {author} {\bibfnamefont {T.}~\bibnamefont {Watashige}}, \bibinfo {author} {\bibfnamefont {Y.}~\bibnamefont {Tsutsumi}}, \bibinfo {author} {\bibfnamefont {T.}~\bibnamefont {Hanaguri}}, \bibinfo {author} {\bibfnamefont {Y.}~\bibnamefont {Kohsaka}}, \bibinfo {author} {\bibfnamefont {S.}~\bibnamefont {Kasahara}}, \bibinfo {author} {\bibfnamefont {A.}~\bibnamefont {Furusaki}}, \bibinfo {author} {\bibfnamefont {M.}~\bibnamefont {Sigrist}}, \bibinfo {author} {\bibfnamefont {C.}~\bibnamefont {Meingast}}, \bibinfo {author} {\bibfnamefont {T.}~\bibnamefont {Wolf}}, \bibinfo {author} {\bibfnamefont {H.~v.}\ \bibnamefont {L\"ohneysen}}, \bibinfo {author} {\bibfnamefont {T.}~\bibnamefont {Shibauchi}},\ and\ \bibinfo {author} {\bibfnamefont {Y.}~\bibnamefont {Matsuda}},\ }\bibfield  {title} {\bibinfo {title} {Evidence for time-reversal symmetry breaking of the superconducting state near twin-boundary interfaces in fese revealed by scanning tunneling spectroscopy},\ }\href
  {https://doi.org/10.1103/PhysRevX.5.031022} {\bibfield  {journal} {\bibinfo  {journal} {Phys. Rev. X}\ }\textbf {\bibinfo {volume} {5}},\ \bibinfo {pages} {031022} (\bibinfo {year} {2015})}\BibitemShut {NoStop}%
\bibitem [{\citenamefont {Sigrist}\ \emph {et~al.}(1996)\citenamefont {Sigrist}, \citenamefont {Kuboki}, \citenamefont {Lee}, \citenamefont {Millis},\ and\ \citenamefont {Rice}}]{sigrist1996influence}%
  \BibitemOpen
  \bibfield  {author} {\bibinfo {author} {\bibfnamefont {M.}~\bibnamefont {Sigrist}}, \bibinfo {author} {\bibfnamefont {K.}~\bibnamefont {Kuboki}}, \bibinfo {author} {\bibfnamefont {P.}~\bibnamefont {Lee}}, \bibinfo {author} {\bibfnamefont {A.}~\bibnamefont {Millis}},\ and\ \bibinfo {author} {\bibfnamefont {T.}~\bibnamefont {Rice}},\ }\bibfield  {title} {\bibinfo {title} {Influence of twin boundaries on josephson junctions between high-temperature and conventional superconductors},\ }\href@noop {} {\bibfield  {journal} {\bibinfo  {journal} {Physical Review B}\ }\textbf {\bibinfo {volume} {53}},\ \bibinfo {pages} {2835} (\bibinfo {year} {1996})}\BibitemShut {NoStop}%
\bibitem [{\citenamefont {Walker}\ and\ \citenamefont {Luettmer-Strathmann}(1996)}]{walker1996josephson}%
  \BibitemOpen
  \bibfield  {author} {\bibinfo {author} {\bibfnamefont {M.}~\bibnamefont {Walker}}\ and\ \bibinfo {author} {\bibfnamefont {J.}~\bibnamefont {Luettmer-Strathmann}},\ }\bibfield  {title} {\bibinfo {title} {Josephson tunneling in high-t c superconductors},\ }\href@noop {} {\bibfield  {journal} {\bibinfo  {journal} {Physical Review B}\ }\textbf {\bibinfo {volume} {54}},\ \bibinfo {pages} {588} (\bibinfo {year} {1996})}\BibitemShut {NoStop}%
\bibitem [{\citenamefont {Feder}\ \emph {et~al.}(1997)\citenamefont {Feder}, \citenamefont {Beardsall}, \citenamefont {Berlinsky},\ and\ \citenamefont {Kallin}}]{feder1997twin}%
  \BibitemOpen
  \bibfield  {author} {\bibinfo {author} {\bibfnamefont {D.}~\bibnamefont {Feder}}, \bibinfo {author} {\bibfnamefont {A.}~\bibnamefont {Beardsall}}, \bibinfo {author} {\bibfnamefont {A.}~\bibnamefont {Berlinsky}},\ and\ \bibinfo {author} {\bibfnamefont {C.}~\bibnamefont {Kallin}},\ }\bibfield  {title} {\bibinfo {title} {Twin boundaries in d-wave superconductors},\ }\href@noop {} {\bibfield  {journal} {\bibinfo  {journal} {Physical Review B}\ }\textbf {\bibinfo {volume} {56}},\ \bibinfo {pages} {R5751} (\bibinfo {year} {1997})}\BibitemShut {NoStop}%
\bibitem [{\citenamefont {Khlyustikov}\ and\ \citenamefont {Moskvin}(1985)}]{khlyustikov1985superconductivity}%
  \BibitemOpen
  \bibfield  {author} {\bibinfo {author} {\bibfnamefont {I.}~\bibnamefont {Khlyustikov}}\ and\ \bibinfo {author} {\bibfnamefont {S.}~\bibnamefont {Moskvin}},\ }\bibfield  {title} {\bibinfo {title} {Superconductivity of the twinning plane of niobium and the topological phase transition in a two-dimensional superconducting system},\ }\href@noop {} {\bibfield  {journal} {\bibinfo  {journal} {Zh. Eksp. Teor. Fiz.}\ }\textbf {\bibinfo {volume} {89}},\ \bibinfo {pages} {1846} (\bibinfo {year} {1985})}\BibitemShut {NoStop}%
\bibitem [{\citenamefont {Khlyustikov}\ and\ \citenamefont {Buzdin}(1987)}]{khlyustikov1987twinning}%
  \BibitemOpen
  \bibfield  {author} {\bibinfo {author} {\bibfnamefont {I.}~\bibnamefont {Khlyustikov}}\ and\ \bibinfo {author} {\bibfnamefont {A.}~\bibnamefont {Buzdin}},\ }\bibfield  {title} {\bibinfo {title} {Twinning-plane superconductivity},\ }\href@noop {} {\bibfield  {journal} {\bibinfo  {journal} {Advances in Physics}\ }\textbf {\bibinfo {volume} {36}},\ \bibinfo {pages} {271} (\bibinfo {year} {1987})}\BibitemShut {NoStop}%
\bibitem [{\citenamefont {Bobrov}\ and\ \citenamefont {Zorin}(1984)}]{bobrov1984increase}%
  \BibitemOpen
  \bibfield  {author} {\bibinfo {author} {\bibfnamefont {V.}~\bibnamefont {Bobrov}}\ and\ \bibinfo {author} {\bibfnamefont {S.}~\bibnamefont {Zorin}},\ }\bibfield  {title} {\bibinfo {title} {Increase of the critical temperature due to low-temperature twinning of superconducting niobium},\ }\href@noop {} {\bibfield  {journal} {\bibinfo  {journal} {JETP Lett.(Engl. Transl.);(United States)}\ }\textbf {\bibinfo {volume} {40}} (\bibinfo {year} {1984})}\BibitemShut {NoStop}%
\bibitem [{\citenamefont {Samoilenka}\ and\ \citenamefont {Babaev}(2020)}]{samoilenka2020boundary}%
  \BibitemOpen
  \bibfield  {author} {\bibinfo {author} {\bibfnamefont {A.}~\bibnamefont {Samoilenka}}\ and\ \bibinfo {author} {\bibfnamefont {E.}~\bibnamefont {Babaev}},\ }\bibfield  {title} {\bibinfo {title} {Boundary states with elevated critical temperatures in bardeen-cooper-schrieffer superconductors},\ }\href@noop {} {\bibfield  {journal} {\bibinfo  {journal} {Physical Review B}\ }\textbf {\bibinfo {volume} {101}},\ \bibinfo {pages} {134512} (\bibinfo {year} {2020})}\BibitemShut {NoStop}%
\bibitem [{\citenamefont {Samoilenka}\ \emph {et~al.}(2020)\citenamefont {Samoilenka}, \citenamefont {Barkman}, \citenamefont {Benfenati},\ and\ \citenamefont {Babaev}}]{samoilenka2020pair}%
  \BibitemOpen
  \bibfield  {author} {\bibinfo {author} {\bibfnamefont {A.}~\bibnamefont {Samoilenka}}, \bibinfo {author} {\bibfnamefont {M.}~\bibnamefont {Barkman}}, \bibinfo {author} {\bibfnamefont {A.}~\bibnamefont {Benfenati}},\ and\ \bibinfo {author} {\bibfnamefont {E.}~\bibnamefont {Babaev}},\ }\bibfield  {title} {\bibinfo {title} {Pair-density-wave superconductivity of faces, edges, and vertices in systems with imbalanced fermions},\ }\href@noop {} {\bibfield  {journal} {\bibinfo  {journal} {Physical Review B}\ }\textbf {\bibinfo {volume} {101}},\ \bibinfo {pages} {054506} (\bibinfo {year} {2020})}\BibitemShut {NoStop}%
\bibitem [{\citenamefont {Benfenati}\ \emph {et~al.}(2021)\citenamefont {Benfenati}, \citenamefont {Samoilenka},\ and\ \citenamefont {Babaev}}]{benfenati2021boundary}%
  \BibitemOpen
  \bibfield  {author} {\bibinfo {author} {\bibfnamefont {A.}~\bibnamefont {Benfenati}}, \bibinfo {author} {\bibfnamefont {A.}~\bibnamefont {Samoilenka}},\ and\ \bibinfo {author} {\bibfnamefont {E.}~\bibnamefont {Babaev}},\ }\bibfield  {title} {\bibinfo {title} {Boundary effects in two-band superconductors},\ }\href@noop {} {\bibfield  {journal} {\bibinfo  {journal} {Physical Review B}\ }\textbf {\bibinfo {volume} {103}},\ \bibinfo {pages} {144512} (\bibinfo {year} {2021})}\BibitemShut {NoStop}%
\bibitem [{\citenamefont {Barkman}\ \emph {et~al.}(2022)\citenamefont {Barkman}, \citenamefont {Samoilenka}, \citenamefont {Benfenati},\ and\ \citenamefont {Babaev}}]{barkman2022elevated}%
  \BibitemOpen
  \bibfield  {author} {\bibinfo {author} {\bibfnamefont {M.}~\bibnamefont {Barkman}}, \bibinfo {author} {\bibfnamefont {A.}~\bibnamefont {Samoilenka}}, \bibinfo {author} {\bibfnamefont {A.}~\bibnamefont {Benfenati}},\ and\ \bibinfo {author} {\bibfnamefont {E.}~\bibnamefont {Babaev}},\ }\bibfield  {title} {\bibinfo {title} {Elevated critical temperature at bcs superconductor--band insulator interfaces},\ }\href@noop {} {\bibfield  {journal} {\bibinfo  {journal} {Physical review b}\ }\textbf {\bibinfo {volume} {105}},\ \bibinfo {pages} {224518} (\bibinfo {year} {2022})}\BibitemShut {NoStop}%
\bibitem [{\citenamefont {Talkachov}\ \emph {et~al.}(2023)\citenamefont {Talkachov}, \citenamefont {Samoilenka},\ and\ \citenamefont {Babaev}}]{talkachov2023microscopic}%
  \BibitemOpen
  \bibfield  {author} {\bibinfo {author} {\bibfnamefont {A.}~\bibnamefont {Talkachov}}, \bibinfo {author} {\bibfnamefont {A.}~\bibnamefont {Samoilenka}},\ and\ \bibinfo {author} {\bibfnamefont {E.}~\bibnamefont {Babaev}},\ }\bibfield  {title} {\bibinfo {title} {Microscopic study of boundary superconducting states on a honeycomb lattice},\ }\href@noop {} {\bibfield  {journal} {\bibinfo  {journal} {Physical Review B}\ }\textbf {\bibinfo {volume} {108}},\ \bibinfo {pages} {134507} (\bibinfo {year} {2023})}\BibitemShut {NoStop}%
\bibitem [{\citenamefont {Hainzl}\ \emph {et~al.}(2023)\citenamefont {Hainzl}, \citenamefont {Roos},\ and\ \citenamefont {Seiringer}}]{hainzl2023boundary}%
  \BibitemOpen
  \bibfield  {author} {\bibinfo {author} {\bibfnamefont {C.}~\bibnamefont {Hainzl}}, \bibinfo {author} {\bibfnamefont {B.}~\bibnamefont {Roos}},\ and\ \bibinfo {author} {\bibfnamefont {R.}~\bibnamefont {Seiringer}},\ }\bibfield  {title} {\bibinfo {title} {Boundary superconductivity in the bcs model},\ }\href@noop {} {\bibfield  {journal} {\bibinfo  {journal} {Journal of spectral theory}\ }\textbf {\bibinfo {volume} {12}},\ \bibinfo {pages} {1507} (\bibinfo {year} {2023})}\BibitemShut {NoStop}%
\bibitem [{\citenamefont {Roos}\ and\ \citenamefont {Seiringer}(2023)}]{roos2023bcs}%
  \BibitemOpen
  \bibfield  {author} {\bibinfo {author} {\bibfnamefont {B.}~\bibnamefont {Roos}}\ and\ \bibinfo {author} {\bibfnamefont {R.}~\bibnamefont {Seiringer}},\ }\bibfield  {title} {\bibinfo {title} {Bcs critical temperature on half-spaces},\ }\href@noop {} {\bibfield  {journal} {\bibinfo  {journal} {arXiv preprint arXiv:2306.05824}\ } (\bibinfo {year} {2023})}\BibitemShut {NoStop}%
\bibitem [{\citenamefont {Roos}(2023)}]{roos2023boundary}%
  \BibitemOpen
  \bibfield  {author} {\bibinfo {author} {\bibfnamefont {B.}~\bibnamefont {Roos}},\ }\emph {\bibinfo {title} {Boundary superconductivity in BCS theory}},\ \href@noop {} {Ph.D. thesis},\ \bibinfo  {school} {Institute of Science and Technology Austria} (\bibinfo {year} {2023})\BibitemShut {NoStop}%
\bibitem [{\citenamefont {Khlyustikov}(1989)}]{khlyustikov1989influence}%
  \BibitemOpen
  \bibfield  {author} {\bibinfo {author} {\bibfnamefont {I.}~\bibnamefont {Khlyustikov}},\ }\bibfield  {title} {\bibinfo {title} {Influence of defects on the critical parameters of twinning plane superconductivity},\ }\href@noop {} {\bibfield  {journal} {\bibinfo  {journal} {Zh. {\'E}ksp. Teor. Fiz}\ }\textbf {\bibinfo {volume} {96}},\ \bibinfo {pages} {2073} (\bibinfo {year} {1989})}\BibitemShut {NoStop}%
\bibitem [{\citenamefont {Song}\ \emph {et~al.}(2012{\natexlab{b}})\citenamefont {Song}, \citenamefont {Wang}, \citenamefont {Jiang}, \citenamefont {Wang}, \citenamefont {He}, \citenamefont {Chen}, \citenamefont {Hoffman}, \citenamefont {Ma},\ and\ \citenamefont {Xue}}]{song2012suppression}%
  \BibitemOpen
  \bibfield  {author} {\bibinfo {author} {\bibfnamefont {C.-L.}\ \bibnamefont {Song}}, \bibinfo {author} {\bibfnamefont {Y.-L.}\ \bibnamefont {Wang}}, \bibinfo {author} {\bibfnamefont {Y.-P.}\ \bibnamefont {Jiang}}, \bibinfo {author} {\bibfnamefont {L.}~\bibnamefont {Wang}}, \bibinfo {author} {\bibfnamefont {K.}~\bibnamefont {He}}, \bibinfo {author} {\bibfnamefont {X.}~\bibnamefont {Chen}}, \bibinfo {author} {\bibfnamefont {J.~E.}\ \bibnamefont {Hoffman}}, \bibinfo {author} {\bibfnamefont {X.-C.}\ \bibnamefont {Ma}},\ and\ \bibinfo {author} {\bibfnamefont {Q.-K.}\ \bibnamefont {Xue}},\ }\bibfield  {title} {\bibinfo {title} {Suppression of superconductivity by twin boundaries in fese},\ }\href@noop {} {\bibfield  {journal} {\bibinfo  {journal} {Physical Review Letters}\ }\textbf {\bibinfo {volume} {109}},\ \bibinfo {pages} {137004} (\bibinfo {year} {2012}{\natexlab{b}})}\BibitemShut {NoStop}%
\bibitem [{\citenamefont {Kirtley}\ \emph {et~al.}(2010)\citenamefont {Kirtley}, \citenamefont {Kalisky}, \citenamefont {Luan},\ and\ \citenamefont {Moler}}]{kirtley2010meissner}%
  \BibitemOpen
  \bibfield  {author} {\bibinfo {author} {\bibfnamefont {J.~R.}\ \bibnamefont {Kirtley}}, \bibinfo {author} {\bibfnamefont {B.}~\bibnamefont {Kalisky}}, \bibinfo {author} {\bibfnamefont {L.}~\bibnamefont {Luan}},\ and\ \bibinfo {author} {\bibfnamefont {K.~A.}\ \bibnamefont {Moler}},\ }\bibfield  {title} {\bibinfo {title} {Meissner response of a bulk superconductor with an embedded sheet of reduced penetration depth},\ }\href@noop {} {\bibfield  {journal} {\bibinfo  {journal} {Physical Review B—Condensed Matter and Materials Physics}\ }\textbf {\bibinfo {volume} {81}},\ \bibinfo {pages} {184514} (\bibinfo {year} {2010})}\BibitemShut {NoStop}%
\bibitem [{\citenamefont {Kalisky}\ \emph {et~al.}(2011)\citenamefont {Kalisky}, \citenamefont {Kirtley}, \citenamefont {Analytis}, \citenamefont {Chu}, \citenamefont {Fisher},\ and\ \citenamefont {Moler}}]{kalisky2011behavior}%
  \BibitemOpen
  \bibfield  {author} {\bibinfo {author} {\bibfnamefont {B.}~\bibnamefont {Kalisky}}, \bibinfo {author} {\bibfnamefont {J.}~\bibnamefont {Kirtley}}, \bibinfo {author} {\bibfnamefont {J.}~\bibnamefont {Analytis}}, \bibinfo {author} {\bibfnamefont {J.-H.}\ \bibnamefont {Chu}}, \bibinfo {author} {\bibfnamefont {I.}~\bibnamefont {Fisher}},\ and\ \bibinfo {author} {\bibfnamefont {K.~A.}\ \bibnamefont {Moler}},\ }\bibfield  {title} {\bibinfo {title} {Behavior of vortices near twin boundaries in underdoped ba (fe 1-x co x) 2 as 2},\ }\href@noop {} {\bibfield  {journal} {\bibinfo  {journal} {Physical Review B—Condensed Matter and Materials Physics}\ }\textbf {\bibinfo {volume} {83}},\ \bibinfo {pages} {064511} (\bibinfo {year} {2011})}\BibitemShut {NoStop}%
\bibitem [{\citenamefont {Noad}\ \emph {et~al.}(2016)\citenamefont {Noad}, \citenamefont {Spanton}, \citenamefont {Nowack}, \citenamefont {Inoue}, \citenamefont {Kim}, \citenamefont {Merz}, \citenamefont {Bell}, \citenamefont {Hikita}, \citenamefont {Xu}, \citenamefont {Liu} \emph {et~al.}}]{noad2016variation}%
  \BibitemOpen
  \bibfield  {author} {\bibinfo {author} {\bibfnamefont {H.}~\bibnamefont {Noad}}, \bibinfo {author} {\bibfnamefont {E.~M.}\ \bibnamefont {Spanton}}, \bibinfo {author} {\bibfnamefont {K.~C.}\ \bibnamefont {Nowack}}, \bibinfo {author} {\bibfnamefont {H.}~\bibnamefont {Inoue}}, \bibinfo {author} {\bibfnamefont {M.}~\bibnamefont {Kim}}, \bibinfo {author} {\bibfnamefont {T.~A.}\ \bibnamefont {Merz}}, \bibinfo {author} {\bibfnamefont {C.}~\bibnamefont {Bell}}, \bibinfo {author} {\bibfnamefont {Y.}~\bibnamefont {Hikita}}, \bibinfo {author} {\bibfnamefont {R.}~\bibnamefont {Xu}}, \bibinfo {author} {\bibfnamefont {W.}~\bibnamefont {Liu}}, \emph {et~al.},\ }\bibfield  {title} {\bibinfo {title} {Variation in superconducting transition temperature due to tetragonal domains in two-dimensionally doped srtio 3},\ }\href@noop {} {\bibfield  {journal} {\bibinfo  {journal} {Physical Review B}\ }\textbf {\bibinfo {volume} {94}},\ \bibinfo {pages} {174516} (\bibinfo {year} {2016})}\BibitemShut {NoStop}%
\bibitem [{\citenamefont {de~Gennes}(1964)}]{de1964boundary}%
  \BibitemOpen
  \bibfield  {author} {\bibinfo {author} {\bibfnamefont {P.}~\bibnamefont {de~Gennes}},\ }\bibfield  {title} {\bibinfo {title} {Boundary effects in superconductors},\ }\href@noop {} {\bibfield  {journal} {\bibinfo  {journal} {Reviews of Modern Physics}\ }\textbf {\bibinfo {volume} {36}},\ \bibinfo {pages} {225} (\bibinfo {year} {1964})}\BibitemShut {NoStop}%
\bibitem [{\citenamefont {Samoilenka}\ and\ \citenamefont {Babaev}(2021)}]{samoilenka2021microscopic}%
  \BibitemOpen
  \bibfield  {author} {\bibinfo {author} {\bibfnamefont {A.}~\bibnamefont {Samoilenka}}\ and\ \bibinfo {author} {\bibfnamefont {E.}~\bibnamefont {Babaev}},\ }\bibfield  {title} {\bibinfo {title} {Microscopic derivation of superconductor-insulator boundary conditions for ginzburg-landau theory revisited: Enhanced superconductivity at boundaries with and without magnetic field},\ }\href@noop {} {\bibfield  {journal} {\bibinfo  {journal} {Physical Review B}\ }\textbf {\bibinfo {volume} {103}},\ \bibinfo {pages} {224516} (\bibinfo {year} {2021})}\BibitemShut {NoStop}%
\bibitem [{\citenamefont {Fulde}\ and\ \citenamefont {Ferrell}(1964)}]{fulde1964superconductivity}%
  \BibitemOpen
  \bibfield  {author} {\bibinfo {author} {\bibfnamefont {P.}~\bibnamefont {Fulde}}\ and\ \bibinfo {author} {\bibfnamefont {R.~A.}\ \bibnamefont {Ferrell}},\ }\bibfield  {title} {\bibinfo {title} {Superconductivity in a strong spin-exchange field},\ }\href@noop {} {\bibfield  {journal} {\bibinfo  {journal} {Physical Review}\ }\textbf {\bibinfo {volume} {135}},\ \bibinfo {pages} {A550} (\bibinfo {year} {1964})}\BibitemShut {NoStop}%
\bibitem [{\citenamefont {Larkin}\ and\ \citenamefont {Ovchinnikov}(1965)}]{larkin1965zh}%
  \BibitemOpen
  \bibfield  {author} {\bibinfo {author} {\bibfnamefont {A.}~\bibnamefont {Larkin}}\ and\ \bibinfo {author} {\bibfnamefont {Y.~N.}\ \bibnamefont {Ovchinnikov}},\ }\bibfield  {title} {\bibinfo {title} {Zh. {\'e} ksp. teor. fiz. 47, 1136 1964 sov. phys},\ }\href@noop {} {\bibfield  {journal} {\bibinfo  {journal} {JETP}\ }\textbf {\bibinfo {volume} {20}},\ \bibinfo {pages} {762} (\bibinfo {year} {1965})}\BibitemShut {NoStop}%
\bibitem [{\citenamefont {Zhitomirsky}\ and\ \citenamefont {Walker}(1997)}]{zhitomirsky1997electronic}%
  \BibitemOpen
  \bibfield  {author} {\bibinfo {author} {\bibfnamefont {M.}~\bibnamefont {Zhitomirsky}}\ and\ \bibinfo {author} {\bibfnamefont {M.}~\bibnamefont {Walker}},\ }\bibfield  {title} {\bibinfo {title} {Electronic states on a twin boundary of a d-wave superconductor},\ }\href@noop {} {\bibfield  {journal} {\bibinfo  {journal} {Physical review letters}\ }\textbf {\bibinfo {volume} {79}},\ \bibinfo {pages} {1734} (\bibinfo {year} {1997})}\BibitemShut {NoStop}%
\bibitem [{\citenamefont {Samoilenka}(2023)}]{SamoilenkaPhD}%
  \BibitemOpen
  \bibfield  {author} {\bibinfo {author} {\bibfnamefont {A.}~\bibnamefont {Samoilenka}},\ }\emph {\bibinfo {title} {Novel Phenomena in Superconductors and Superfluids: Boundary States, Spiral Magnetic Fields, and Solitons}},\ \href@noop {} {Ph.D. thesis},\ \bibinfo  {school} {KTH, Condensed Matter Theory} (\bibinfo {year} {2023}),\ \bibinfo {note} {qC 2023-04-25}\BibitemShut {NoStop}%
\bibitem [{\citenamefont {Suslov}(1989)}]{suslov1989mechanism}%
  \BibitemOpen
  \bibfield  {author} {\bibinfo {author} {\bibfnamefont {I.}~\bibnamefont {Suslov}},\ }\bibfield  {title} {\bibinfo {title} {Mechanism of the superconductivity of twinning planes},\ }\href@noop {} {\bibfield  {journal} {\bibinfo  {journal} {Zh. Eksp. Teor. Fiz}\ }\textbf {\bibinfo {volume} {95}},\ \bibinfo {pages} {965} (\bibinfo {year} {1989})}\BibitemShut {NoStop}%
\end{thebibliography}%

\end{document}